\documentclass[11pt,twoside]{article}

\usepackage{asp2006}
\usepackage{epsf}
\usepackage{epsfig}
\usepackage{lscape}
\begin{document}
\title{HW Vir's Companion: a M-type Dwarf, or maybe a giant 
  rotating spherical Mirror?}   
\author{H.~Edelmann}   
\affil{The University of Texas at Austin, McDonald Observatory,
1 University Station C1402, Austin, TX 78712, USA}    

\begin{abstract} 
%
From optical high-resolution spectra the nature of
the unseen companion of HW~Vir is determined
without detection of any spectral features originating 
from the secondary itself.
Using radial velocity measurements from the primary hot
subdwarf B star and from weak additional absorption lines 
detected close 
to the secondary eclipse, probably caused by reflected light
off the surface of the secondary, the mass and 
radius of the companion is determined. The values are 
consistent with those of a M type main sequence star.
%
\end{abstract}


\section{Introduction}   
HW~Vir ($\alpha_{2000}=12^{\rm h} 44^{\rm m} 20\fs{2},\, 
\delta_{2000}=-08\deg 40\arcmin 16\farcs{8}$)
is a well analyzed bright ($B=10\fm5$) object.
Berger \& Fringant \citeyearpar{ber80} were the first to 
classify it 
as a subdwarf B (sdB) star using
$UV$ low resolution spectroscopy. 
From photometry Menzies \citeyearpar{men86} discovered that
HW~Vir is not a single star but an eclipsing
binary  with an orbital period of $P=2^{\rm h}48^{\rm m}$
showing a strong reflection effect.  
Subsequent radial velocity measurements (Menzies \& Marang 
\citeyear{mma86}) revealed a first determination of the semi amplitude 
of $K_{\rm sdB} = 87.9 \pm
4.8~{\rm km\, s^{-1}}$ for the primary hot subdwarf
and a system velocity of $\gamma_0=-9.1\pm0.9~{\rm km\, s^{-1}}$, 
but from their 
spectra they could not detect any lines due to the secondary.   
Using \ubvr\ photometry Wood, Zang, \& Robinson \citeyearpar{wzr93}
determined the inclination angle of the system to
$i=80\fdg{6}\pm 0\fdg{2}$. 
Independently, Wood, Zang, \& Robinson \citeyearpar{wzr93} as well as 
W{\l}odarczyk \& Olszewski \citeyearpar{wlo94}, who used $BVR$ photometry,
deduced from their data that the primary star
is most likely a M-type dwarf with a mass near $0.15~M_\odot$.
In \citeyear{woo99} Wood \& Saffer presented a paper
about the analysis of more than two dozen of optical
low resolution spectra taken from HW~Vir in order to measure
not only the orbital parameters of the system 
($K_{\rm sdB}=82.3\pm4.0~{\rm km\, s^{-1}}$,
$\gamma_0=+2.9\pm3.1~{\rm km\, s^{-1}}$)
but also
to determine the atmospheric parameters of the sdB star.
Their results show that the primary is a bona-fide 
sdB star ($T_{\rm eff}=28\,488\pm208~{\rm K}$,
$\log(g)=5.63\pm0.03$,
$N({\rm He})/n({\rm H})=0.0066\pm0.0004$)
at a distance of $D=171 \pm 19$~pc, and the secondary's 
star mass and radius are consistent with those of a M-type dwarf.
But they discovered something else, which was previously unknown:
from their co-added single spectra they were able to detect
a very weak {\it absorption} feature close to the ${\rm H}_\alpha$ line
of the primary sdB star, however, only visible close to the
secondary eclipse. 
Because the radial velocity of this line matched very good the value 
one would expect to find for a M-type dwarf companion,
Wood \& Saffer \citeyearpar{woo99} favored the interpretation that this line 
is a result of an irradiation of the face of the secondary. 
But they could not elucidate why this line is showing in
absorption and not, like expected, in emission.
High resolution observations from Hilditch, Harries, \& Hill
\citeyearpar{hil96} revealed, like all observations before, no indication
of any spectral features from the companion.
However, their spectra only covered the wavelength region  
of 4360-6070~\AA.

Here, I present new results from high resolution spectroscopy
obtained for HW~Vir.
%
\section{Observations and orbital Parameters}   
\label{obs}
A set of high resolution spectra have been obtained 
on the night of January 30-31, 2006 at the McDonald observatory,
West-Texas, USA, with the 2.7~m Harlan J. Smith telescope equipped with the 
high resolution cross-dispersed echelle spectrometer (cs2).
In order to observe one complete cycle ($P=2^{\rm h}48^{\rm m}$)
16 consecutive spectra, with an integration time of 10 minutes each,
were obtained.
The observing time was chosen to 
retrieve a reasonable S/N$\sim$30 on one hand,
but on the other hand also to minimize orbital smearing.
The spectra cover the wavelength from 3\,600~\AA\ to 10\,000~\AA.
The nominal resolution is $\lambda/\Delta\lambda = 35\,000$.

The radial velocities (RVs) for the primary sdB star are determined 
by calculating the shifts of
the measured wavelengths of H$_{\alpha}$, H$_{\beta}$, and 
\ion{He}{i} 5876~\AA. To determine the central wavelengths Gaussian curves 
are fitted to the Balmer line cores and to the helium line.
After the measurement the RVs were corrected to heliocentric values.
Using the same method described by Edelmann et al. \citeyearpar{ede05}
a period of $P=0.115(8) \mbox{ days} = 2^{\rm h} 46^{\rm m} \pm 12^{\rm m}$,
a semi amplitude of $K_{\rm sdB} = 84.6 \pm 1.1~{\rm km\, s^{-1}}$,
and  
a system velocity of $\gamma_0 = -13.0 \pm 0.8~{\rm km\, s^{-1}}$
results.  
Fig. 
\ref{hwvir_phase_teff}  (top pane)
shows the best 
fit RV curve. 
The orbital results are in very good agreement to the parameters
determined by Wood, Zhang, \& Robinson \citeyearpar{wzr93}: 
$P=2^{\rm h} 48^{\rm m}$, 
and Wood \& Saffer \citeyearpar{woo99}:   
$K_{\rm sdB} = 82.3 \pm 4.0~{\rm km\, s^{-1}}$,
and
$\gamma_0 = +2.9 \pm 3.1~{\rm km\, s^{-1}}$.

The residuals to the sine fit plotted in Fig. 
\ref{hwvir_phase_teff} (second panel from the top) 
indicate that the majority of
the RV values are well reproduced by a sinusoidal curve.
Only 2 measurements close to phase zero, i.e. the primary eclipse, 
are 
off the predicted values. The first value is much too high 
(i.e. the emitting source is moving away faster), and the second
value is much too low (i.e. the emitting source is moving faster
towards the observer). 
Lets assume the sdB star is rotating.
Shortly before the total eclipse occurs the hemisphere
of the sdB which is moving towards us is covered by the
surface of the companion, so the measured RV value is the
velocity of the star moving away from us due to the orbital 
motion {\bf plus} the velocity of the hemisphere  
moving away from us due to the rotation. Shortly after the
primary eclipse the measured RV value is the  
velocity of the star moving towards us {\bf plus} the velocity 
of the hemisphere moving towards us. The other hemisphere is, again, 
covered by the companion.
This effect is know as the Rossiter-McLaughlin (RM) effect 
(Rossiter \citeyear{ros24}, McLaughlin \citeyear{mcl24}).
Detecting an RM effect means the sdB star
must rotate significantly, and this should be seen in
the observed line profiles (cf. Section \ref{atmo_para}).
A quantitative investigation to determine the radial velocity
using the RM effect is, however, beyond the scope of this
proceedings paper. It will be addressed in a subsequent, more detailed
paper.
The RM effect also has been reported for other eclipsing binaries
containing a hot subdwarf star (e.g. AA Dor, Rauch \& Werner
\citeyear{rau03}; NY Vir, Vu\v{c}ovi\'{c} et al. \citeyear{vuc07} and this 
proceedings). 
%
\section{Nature of the Companion}
\begin{figure}
\vspace{8.5cm}
\includegraphics{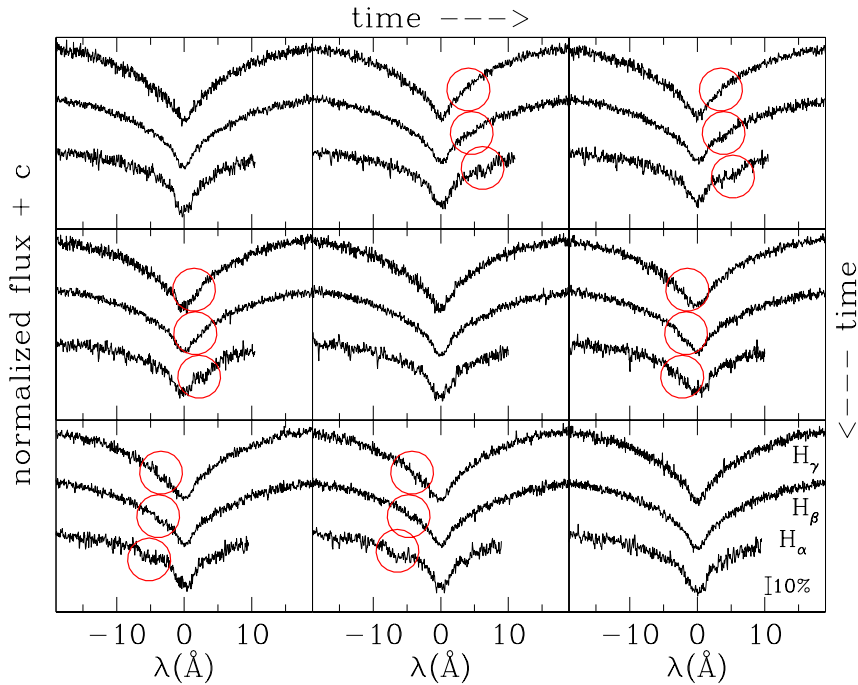}
 \caption[]{Time sequence of 9 consecutive observations showing the 3 mayor 
 balmer lines (each panel from bottom to top: ${\rm H}_\alpha$, ${\rm H}_\beta$,
 and ${\rm H}_\gamma$) for HW~Vir shifted to the rest frame velocity 
 of the sdB star. The center panel shows the observation
 done during the eclipse of the secondary M-type dwarf. 
 Top left hand, center, and bottom right hand panel: only Balmer lines
 originating from the primary sdB star can be seen. All other panels: 
 a set of balmer lines, possibly due to the secondary, can be detected
 moving in time from the red to the blue side.
\label{plot_balmer_2}
}
\end{figure}

%
The high resolution data obtained show at first sight
only single lined
spectra of a RV variable sdB star with an invisible 
companion. However, watching more carefully you can detect,
close to the secondary eclipse 
some sort of 'additional' very weak 
absorption lines in the wings of the sdB's Balmer lines
moving in time from the 
red to the blue hand side (see Fig. \ref{plot_balmer_2}).

The radial velocities that can be measured from each of those lines
in each individual spectra match, i.e. the RV of the line in the wing of 
${\rm H}_\alpha$ match to those of the one in the wing of 
${\rm H}_\beta$, and also to those for ${\rm H}_\gamma$\footnote{The 
secondary absorption lines due to ${\rm H}_\gamma$ 
are barely above the detection limit}.
Therefore, these lines are connected and represent very likely 
a Balmer line series from another object, possibly due to the secondary.
Because you can {\bf only} see the 'secondary' Balmer lines close to
the secondary eclipse, i.e. the cool companion is covered by
sdB star, which in turn also means that in this phase you see
most of the surface of the companion illuminated by the
hot subdwarf, the lines are possibly due, and {\bf solely due}, to the
heated hemisphere of the companion. 
However, then, if assuming the star is rotating,
our RV measurements cannot be 'directly' used to determine 
the RV curve. During 
quadrature\footnote{In quadrature the sdB is moving away 
(or moving towards us) at the same time as the secondary 
is moving towards us (or moving away) at maximum speed.},
i.e. phase=0.25, 0.75, to the observer only one half of the 
secondary's sphere
is illuminated (an therefore heated). This means
that you do not measure  the total RV at this phase, but something
like the RV minus the rotation of the star.
So one can only use the RV values close to the eclipse,
when the rotation of the star does not influence the 
RV measurements, to determine the RV curve.
With the fixed period of $P=0.115~{\rm days}$, and a fixed 
system velocity of $\gamma_0 = +2.9~{\rm km\, s^{-1}}$,
a $K_{\rm sec}=293 \pm 18~{\rm km\, s^{-1}}$ results.
The measured RV values and the fitted curve to the 2 points
close to the secondary eclipse is show in 
Fig. \ref{hwvir_phase_teff} (dotted line).
Using the systems mass function and Kepler's 3rd law
\[
f_m = \frac{M_{\rm comp}^3 \, \sin^3(i)}{(M_{\rm sdB}+M_{\rm comp})^2}
    = \frac{PK_{\rm sdB}^3}{2\pi G} \qquad , \qquad 
       \frac{K_{\rm sdB}}{K_{\rm comp}} = \frac{M_{\rm comp}}{M_{\rm sdB}}
\]
the mass of the sdB primary, as well as those of the companion,
can be calculated to $M_{\rm sdB} = 0.53 \pm 0.08~M_\odot$, and 
$M_{\rm sec} = 0.15 \pm 0.03~M_\odot$.

\begin{figure}
\vspace{8.5cm}
\includegraphics{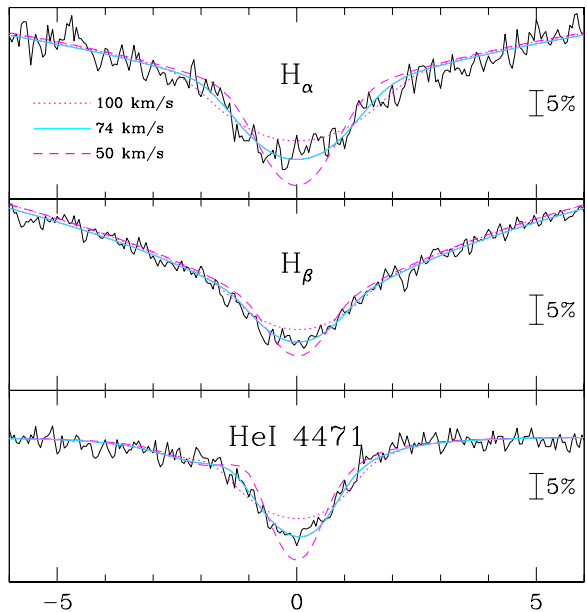}
 \caption[]{Line profiles for ${\rm H}_\alpha$, ${\rm H}_\beta$,
 and \ion{He}{i} 4471~\AA\ calculated for 3 different projected
 rotational velocities of $v\, \sin(i) = 50$, 74, and 100~km/s 
 (dashed, straight,
 and dotted lines, respectively), compared to the observed line profiles 
 of HW~Vir observed close to the primary eclipse.
\label{plot_48_vrad}
}
\end{figure}
The results are very promising, because not only the mass of
the sdB is close to its canonical mass, like the majority of all
measured sdB masses are, but also the mass of the companion is
in perfect agreement with the results from the photometric 
observations (e.g. Wood, Zang, \& Robinson \citeyear{wzr93},
W{\l}odarczyk \& Olszewski \citeyear{wlo94}).

The difference between 
the maximum (minimum) value of the secondary's RV curve
and the measured RV value of the lines close to the
quadrature phase is about $80~{\rm km\, s^{-1}}$. 
As described above, this is
also roughly the rotational velocity of the companion.
Assuming a bound rotation of the secondary 
it is also possible to estimate the radius of the companion to
$R_{\sec} = 
(2 \pi)^{-1} P v_{\rm rot}^{\rm sec} \approx 
0.19~R_\odot$. 
This very rough estimation is in agreement with the results
from the photometric work of Wood, Zang, \& Robinson (\citeyear{wzr93},
$R_{\sec} = 0.188~R_\odot$), and W{\l}odarczyk \& Olszewski (\citeyear{wlo94}
$R_{\sec} = 0.18~R_\odot$). 
The mass and radius determined for HW~Vir's companion is also
consistent with theoretical models for low-mass dwarf stars
from Dorman et al \citeyearpar{dor89}: a $0.15~M_\odot$ star should have 
a radius of $0.18~R_\odot$.
The companion, therefore, is a M type dwarf.
%
\section{Atmospheric Parameters of the sdB Star}
\label{atmo_para}
%
\begin{figure}
\vspace{13cm}
\includegraphics{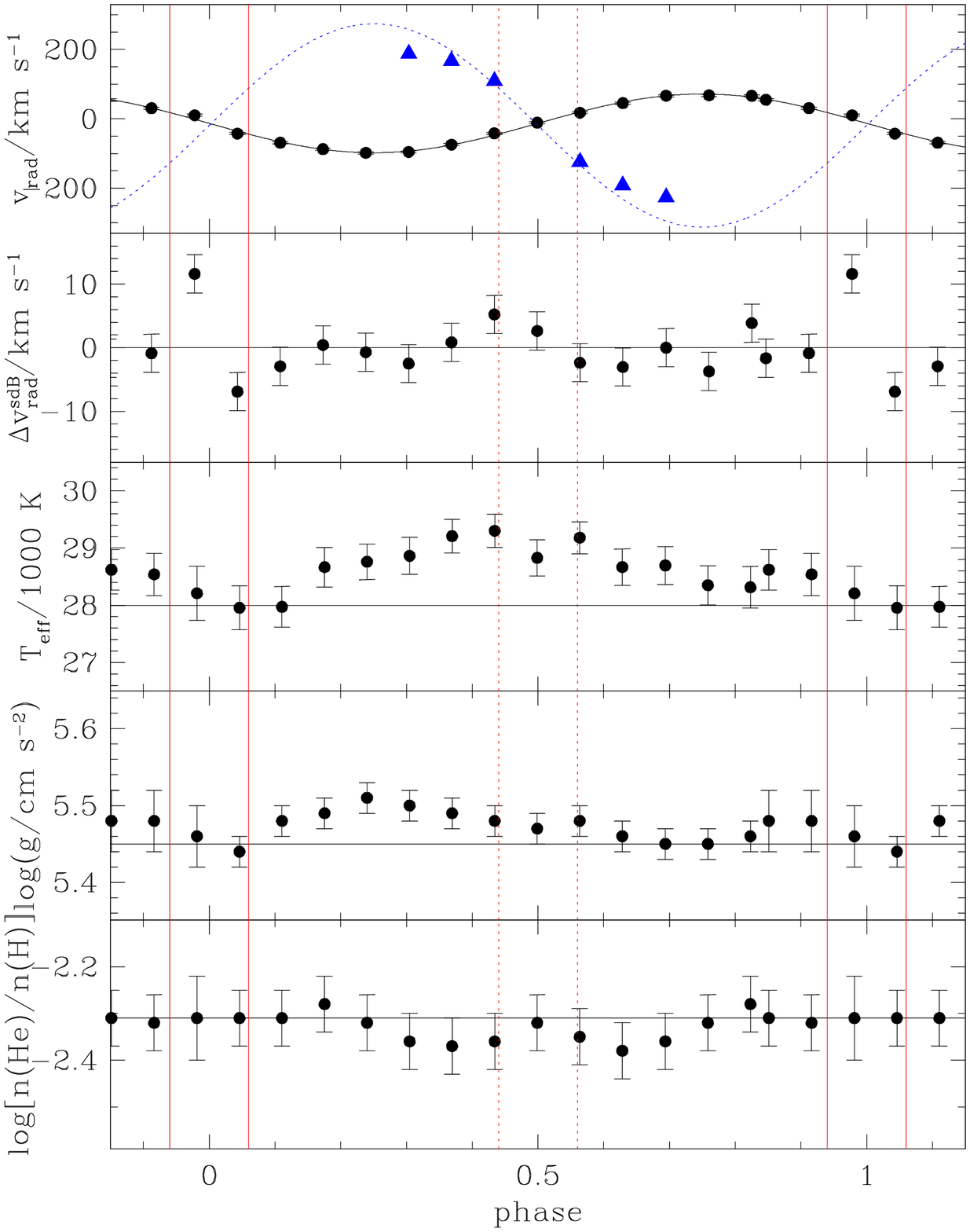}
 \caption[]{{\it Top panel}: Measured radial velocities as 
 a function of orbital phase for the sdB primary (dots) and
 for the M-type dwarf companion (triangles) together with
 a fitted sine curve for all observed radial velocities
 for the sdB primary (solid line), 
 and a sine curve fitted to the 2 points close to
 the secondary eclipse (period and system velocity fixed, dotted line).\\  
 {\it Second panel from the top}: Residuals for the
 sdB primary ($\Delta v_{\rm rad}=v_{\rm obs}-v_{\rm sine}$) 
 to the sine fit including error bars.
 Note the 'outliers' during the primary eclipse (start and end are 
 marked by the vertical solid lines) which are due to the 
 Rossiter-McLaughlin effect (see text).\\
 Also plotted are the determined effective temperature ({\it center panel}),
 surface gravity ({\it second panel from the bottom}), and
 helium abundance ({\it bottom panel}) for all individual 
 spectra including error bars.\\
 The vertical dotted lines mark the start and end of the secondary 
 eclipse.
\label{hwvir_phase_teff}
}
\end{figure}
The stellar atmospheric parameters are determined for all individual
normalized (the continuum was set to unity) spectra
using a $\chi^2$ fit method (Napiwotzki, Green, \& Saffer \citeyear{nap99})
and NLTE model atmospheres from Napiwotzki \citeyearpar{nap97}. 
In this fit procedure I also included the projected rotational velocity
as an additional parameter, because the spectral lines show clear 
indications of rotational broadening (cfg. Section \ref{obs}, RM-effect).
The resulting effective temperature, surface gravity, and 
helium abundance for every spectrum are shown in Fig. \ref{hwvir_phase_teff}.
A projected rotational velocity of 
$v_{\rm rot}^{\rm sdB}=74\pm2~{\rm km\, s^{-1}}$ ($3\sigma$ error) 
results (see Fig. \ref{plot_48_vrad}). 
At first sight the results looks like one would expect from a star in a 
binary showing a tidal deformation. However, the phase does not match to this
image. Also the trend of the effective temperature during the secondary 
eclipse makes no sense. Why should the temperature suddenly drop about 
500~K when the cool companion is, in the line of sight, behind the
hot subdwarf? 

The developing of the temperature curve reminds someone 
extremely to the developing of the light curves measured for HW~Vir
e.g. by Wood, Zhang, \& Robinson (\citeyear{wzr93}, Fig. 2). After the   
primary eclipse the flux is increasing because the visible 
illuminated surface of the companion is getting bigger and bigger, 
and more and more light is reflected towards the observer,
until the secondary eclipse occurs. 
During the secondary eclipse the flux suddenly drops because the sdB star 
covers the illuminated (reflecting) surface of the cool companion. And
after the secondary eclipse the flux is again decreasing, because the
less and less light is reflected from the surface of the secondary
to the observer
because the illuminated surface is getting smaller and smaller.
The same trend can be seen in the measured temperature curve in
Fig. \ref{hwvir_phase_teff}.
After the primary eclipse $T_{\rm eff}$ increases until the secondary eclipse
happens.
Here it suddenly drops, and after the eclipse $T_{\rm eff}$ decreases 
again until the primary eclipse takes place.    
Therefore, the determined atmospheric parameters are possibly
influenced by the reflected light.

\begin{figure}
\vspace{9.0cm}
\includegraphics{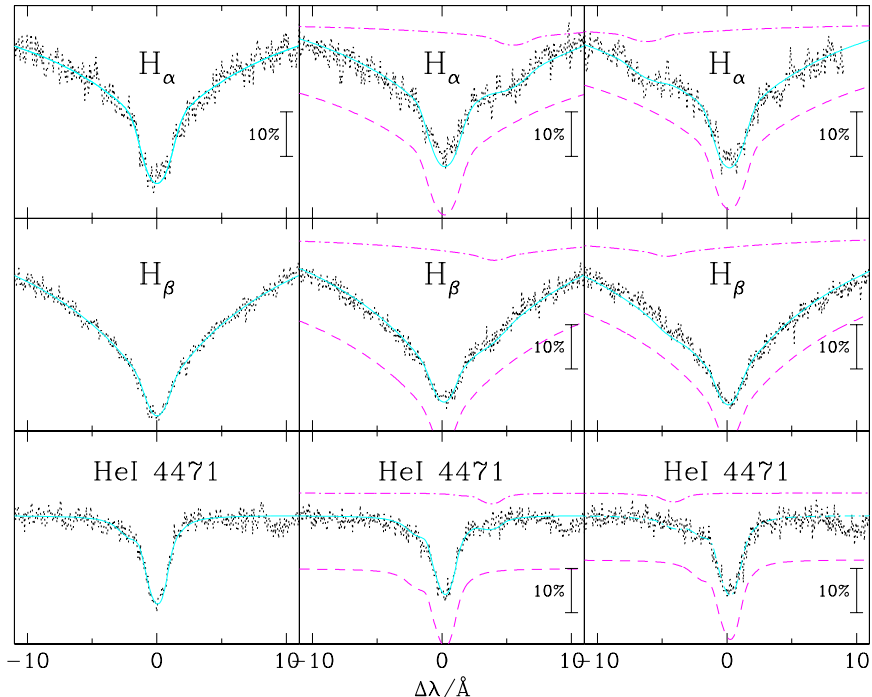}
 \caption[]{{\it Left hand panel}: observed spectrum close to the primary eclipse 
 (dotted lines), and theoretical NLTE line profiles (solid lines) for the
 parameters retrieved from the fits close to the primary eclipse,
 horizontal lines in the
 lower 3 panels).
 {\it Center panel}: observed spectra close before the secondary eclipse
 (dotted lines), 88\% 'direct light' from the sdB star itself
 (dashed lines), 12\% reflection (dashed dotted lines, shifted in
 y-axis for clarity), and sum of both components (solid line)
 using the {\bf same} NLTE model spectrum for the 
 direct light {\bf and} the reflection. 
 {\it Right hand panel}: observed spectra close after the secondary eclipse
 (dotted lines), 90\% 'direct light' from the sdB star itself
 (dashed lines), 10\% reflection (dashed dotted lines, shifted in
 y-axis for clarity), and sum of both components (solid line)
  using the {\bf same} NLTE model spectrum for the 
 direct light {\bf and} the reflection. 
\label{plot_48_53_58}
}
\end{figure}

To test this theory, I calculated 
one NLTE model spectrum
using the atmospheric parameters derived from the fits close to the
primary eclipse: $T_{\rm eff}=28\,000$~K, $\log(g)=5.45$, and 
$\log[n({\rm He})/n({\rm H})]=-2.32$ (cfg. Fig. \ref{hwvir_phase_teff},
horizontal lines in the lower 3 panels), because
this parameters are least influenced by any reflected light
(cfg Fig. \ref{plot_48_53_58}, left hand panel).
%
Fig. \ref{plot_48_53_58} (center and right hand panel) exemplarily shows
for two observed spectra close to the secondary eclipse 
that the observed line profiles can be matched simply by using 
the sum of 2 of such identical model spectra only shifted 
to the corresponding RV values of the sdB and the M dwarf,
and convolved by a constant factor (the sum of both factors 
need to be one, because all spectra have been normalized to
unity), respectively. 
Also all other spectra can be matched using the same method.

I conclude that the varying parameters shown in 
Fig. \ref{hwvir_phase_teff}
are not caused by a tidal deformation of the sdB, but are 
only due to reflected light from the surface of the secondary. 
%
\section{Summary}
%
A set of 16 consecutive optical high-resolution spectra
have been obtained to search for a spectral feature 
from the (so far unseen) companion of HW~Vir. 
Close to the secondary eclipse
a series of additional weak Balmer absorption lines
can be seen
probably due to the heated hemisphere of the secondary.
The secondary's mass of $M_{\rm sec} = 0.15 \pm 0.03~M_\odot$
and radius of $R_{\sec} = 0.19~R_\odot$ 
is consistent with a M type main sequence star.
The mass of the primary sdB star could be
measured to $M_{\rm sdB} = 0.53 \pm 0.08~M_\odot$.
The radial velocity measurements of the sdB star 
show clear indication of a Rossiter-McLaughlin effect.
The projected rotational velocity of the sdB star 
is measured from the observed line profiles to
$v_{\rm rot}^{\rm sdB}=74\pm2~{\rm km\, s^{-1}}$.
The additional weak Balmer absorption lines
as well as 
varying atmospheric parameters determined from
$\chi^2$-fits of the individual spectra are probably 
due to reflected light on the surface of the cool 
M type dwarf
originating from the hot subdwarf itself.
\acknowledgements 
I would like to thank J.~Tomkin, L.~Koesterke, S.~Nessß-linger, U.~Heber,  
and S.~Geier for lots of fruitful discussions. 
I appreciate the help, support, and valuable assistance provided by 
David Doss at the McDonald observatory, during my visit.


\end{document}